\documentclass[a4paper]{VisionStyle}
\usepackage{epsfig}

\newcommand{\msun}{$M_{\odot}$}

\begin{document}

\title{X-ray transient AGN and galaxies, and why we need new soft X-ray surveys}

\author{D.\,Grupe\inst{1}  } 

\institute{
Max-Planck-Institut f\"ur extraterrestrische Physik, Giessenbachstr. D-85748
Garching, Germany}

\maketitle 

\begin{abstract}
X-ray transience is the most extreme form of variability observed in AGN or
normal non-active galaxies. While factors of 2-3 on timescales of days to years
are quite common among AGN, X-ray transients appear only once and vanish from
the X-ray sky years later. The ROSAT All-Sky Survey with its sensitivity to
energies down to 0.1 keV
was the an ideal tool to discover
these sources. X-ray transience in AGN or galaxies can be caused by dramatic
changes in the accretion rate of the central black hole or by changes of the
properties of the accretion disk. So far only a handful of sources are known. In
order to estimate how often such an event occurs in a galaxy, a new soft X-ray
survey is needed. In these proceedings I describe the currently known X-ray
transient AGN and galaxies and will argue for a new soft X-ray survey in order
to discover more of these extreme X-ray sources.

\end{abstract}

\section{Introduction}
The ROSAT All-Sky Survey (RASS, \cite{dgrupe-WB1:vog99}) 
has established a new phenomenon among AGN and
galaxies - X-ray transience. These X-ray transient AGN and galaxies are very
bright once and appear to be fainter or even vanish in X-rays when observed
years later(e.g. \cite{dgrupe-WB1:gr01a}, \cite{dgrupe-WB1:ko99a}).
X-ray transience can appear in Active Galactic Nuclei as well as in normal
non-active galaxies. While the X-ray transience in AGN might be caused by
changes of the accretion disk properties, in non-active galaxies it might be
caused by an X-ray outburst. These X-ray outbursts can be caused by an
accretion event, either by instabilities in the accretion disk or by the
disruption of a star by the central black hole. Disk instabilities may cause an
X-ray outburst in AGN while the tidal disruption model is favoured to explain
the outbursts in non-active galaxies which do not show any signs of nuclear
activity.

So far only in one source a
response in the optical emission lines has been observed, IC 3599. Future soft
X-ray surveys will allow us to search for this type of sources. Fast follow-up
observations in the optical and in X-rays will provide us with a powerful tool
to map the inner region of an AGN in order to locate the line emitting regions
and to describe what the geometry of the whole system is.

\section{The Sample}
\begin{figure}[ht]
  \begin{center}
    \epsfig{file=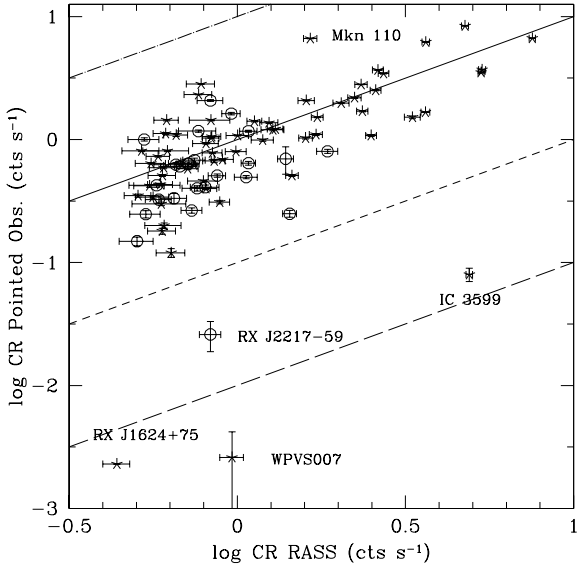, width=8cm}
  \end{center}
\caption{\label{dgrupe_WB1:rass_po}RASS vs. pointed observation count rates 
(taken from Grupe et al. 2001a)
}  

\end{figure}

Our sample of soft X-ray AGN contains 113 sources selected from the RASS by
the PSPC count rates CR $>$ 0.5 cts
s$^{-1}$ and the hardness ratio HR $<$0.0 (\cite{dgrupe-WB1:gru98},
\cite{dgrupe-WB1:gr01a}, \cite{dgrupe-WB1:tho99}).
Pointed PSPC and HRI observations are
available for 60 and 50 sources, respectively. All in all, for more than 80 
sources at least one pointed observation is available (\cite{dgrupe-WB1:gr01a}).
In this way we have a
tool to search for long-term large amplitude variations. 
Fig. \ref{dgrupe_WB1:rass_po} displays the
RASS vs. pointed observation count rates. HRI count rates have been converted
into PSPC count rates assuming no spectral change between both observations.
The solid line marks no change, the short-dashed line a change by a factor of 10
and the long-dashed line by a factor of 100 between the RASS and the pointed
observation. Four sources turned out to vary by factors of almost 100 or even
more: {\bf IC 3599, WPVS007, RX J1624.9+7554}, and {\bf RX J2217.9--5941}. 
The first three are X-ray transients while  RX J2217.9--5941 is an
X-ray transient candidate.

\section{X-ray transient AGN and galaxies}

\subsection{WPVS007}
The Narrow-line Seyfert 1 galaxy (NLS1)
WPVS 007 was {\it the} softest AGN observed during the RASS
(\cite{dgrupe-WB1:gr95b})
 and had a mean count rate of about 1 cts s$^{-1}$. In
later pointings the source shows only the flux expected from a normal non-active
galaxy (\cite{dgrupe-WB1:gr01a}). The flux measured during the RASS observation
was what was expected from an average $f_{\rm X}/f_{\rm opt}$ ratio of an AGN.
The question that has to be asked for WPVS 007 is not why it was bright during
the RASS it is more why it is off now and what caused the dramatic switch off of
this source in less than three years.
A possible scenario to explain this dramatic turn-off is a temperature
change in the Comptonization layer of the accretion disk that shifts the soft
X-ray photons out of the ROSAT energy window. Because no optical variability has
been detected in this source, it can be expected that the AGN engine is still
running and the source should appear in the X-ray sky again. Therefore a
monitoring of this source by XMM or Chandra is needed.

\subsection{IC 3599}
\begin{figure}[ht]
  \begin{center}
    \epsfig{file=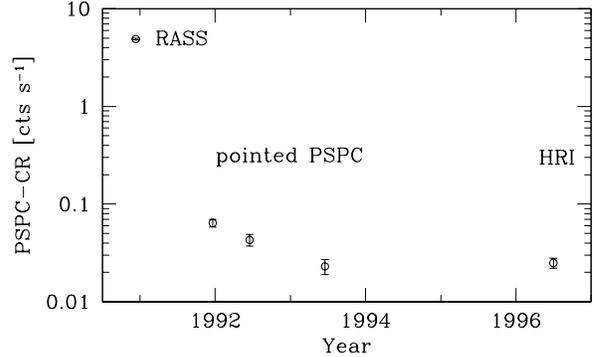,width=8cm,bbllx=1.0cm,bblly=14.8cm,bburx=16.2cm,bbury=24.5cm,clip=}
  \end{center}
\caption{\label{dgrupe_WB1:ic3599_cr} 
Long-term lightcurve of IC 3599 (taken from Grupe et al.
2001a)
}
\end{figure}

\begin{figure*}[ht]
    \epsfig{file=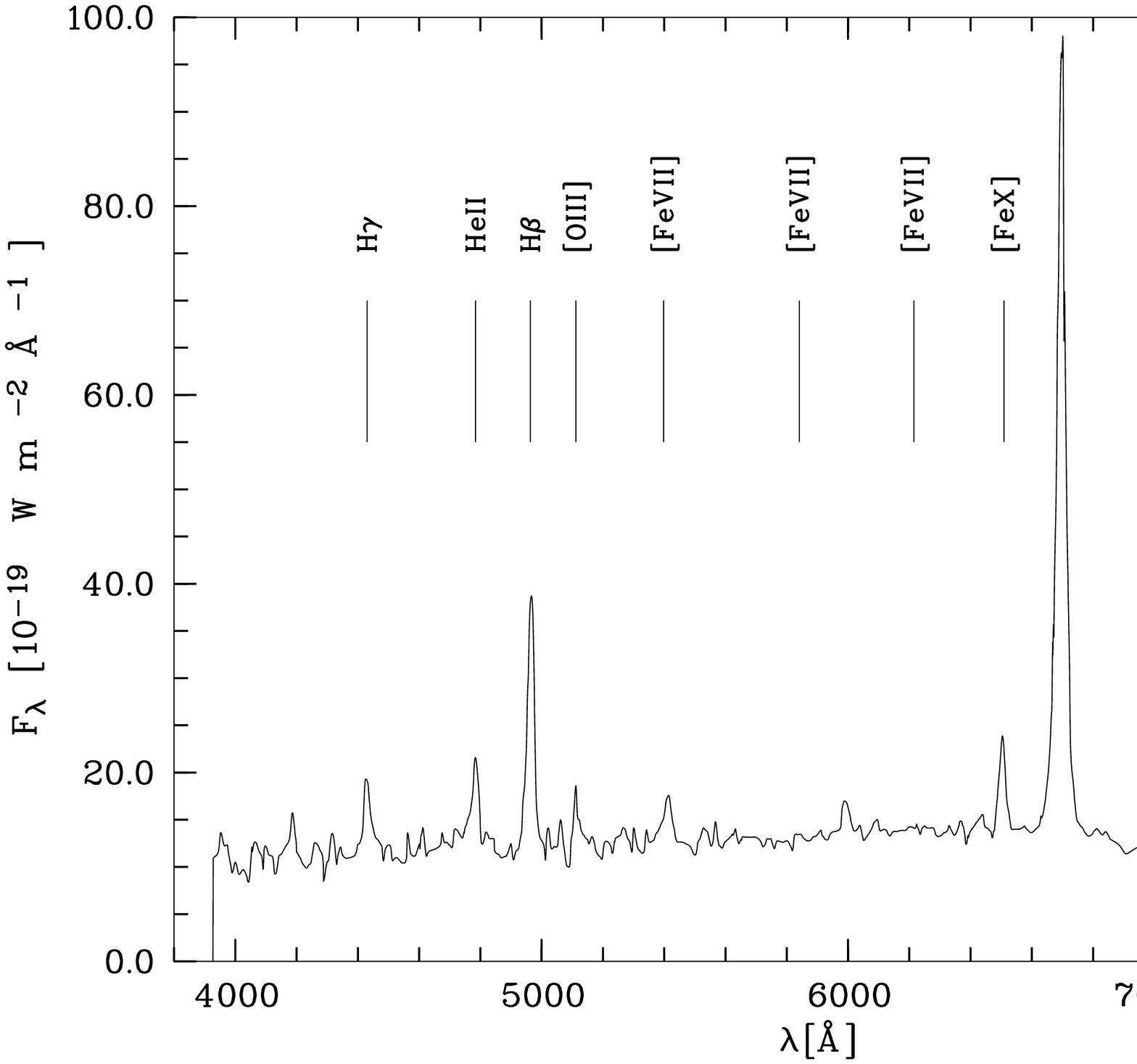,width=8.6cm,bbllx=2.3cm,bblly=1.0cm,bburx=27.5cm,bbury=19.2cm,clip=}
 \hspace*{0.5cm}   
    \epsfig{file=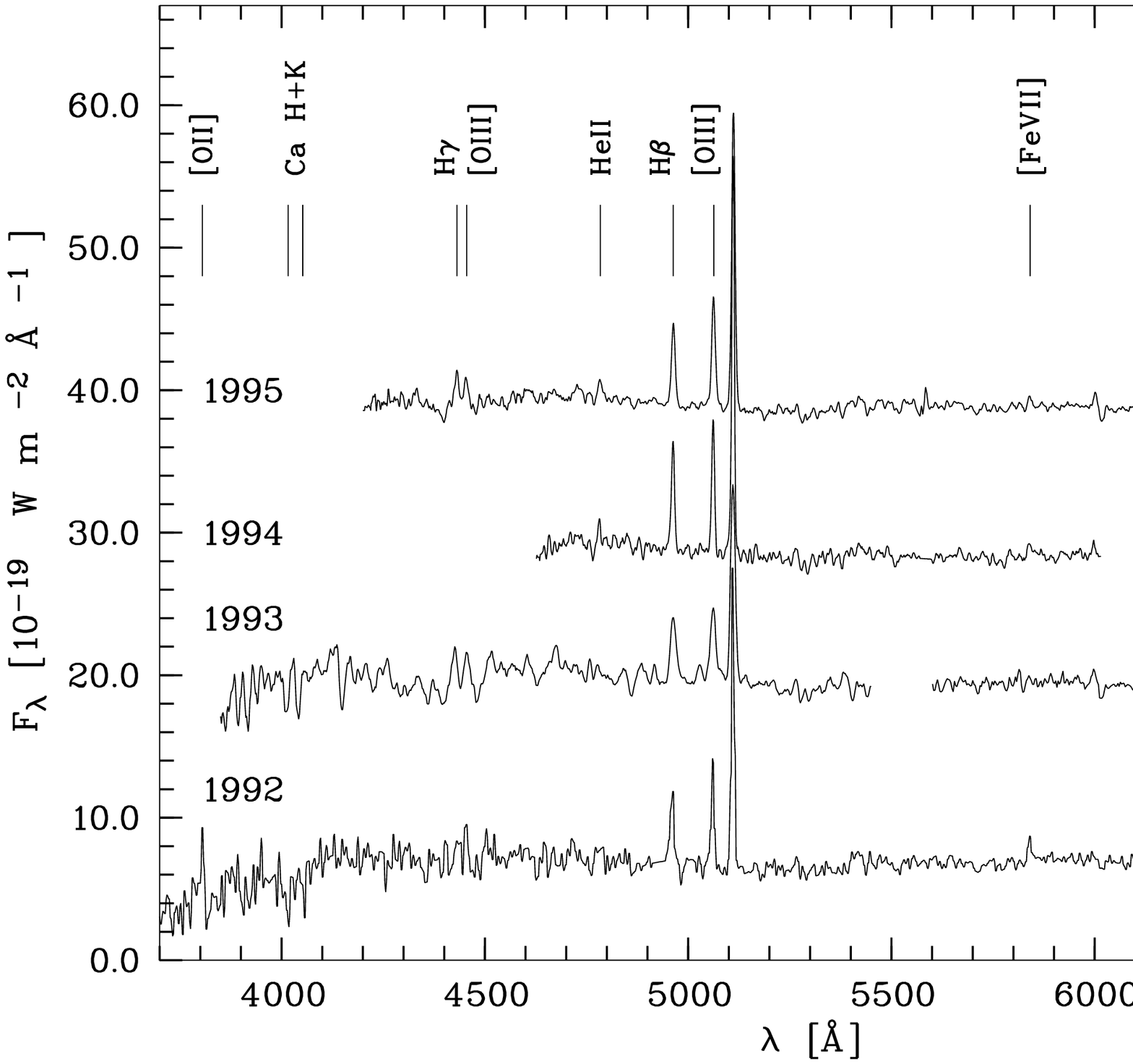, width=8.6cm,bbllx=2.3cm,bblly=1.0cm,bburx=27.5cm,bbury=19.2cm,clip=}
\caption{ \label{dgrupe_WB1:ic3599_opt}
Optical spectra of IC 3599. The left panel displays the spectrum taken
half a year after the X-ray outburst during the RASS (taken from
Brandt et al. 1995) and the right panel shows the spectra taken years later
(Grupe et al. 1995a). The spectra of 1993, 1994, and 1995 are displayed with an
offset.
}  
\end{figure*}

The Seyfert 2 galaxy IC 3599 has shown an X-ray outburst during the RASS
(see Fig. \ref{dgrupe_WB1:ic3599_cr}).
 During the RASS
observation it was one of the brightest AGN in the X-ray sky (4.90 PSPC cts
s$^{-1}$) and has been even seen in ROSAT's Wide Field Camera
(\cite{dgrupe-WB1:pounds93}; \cite{dgrupe-WB1:edel99}). In later pointed PSPC and HRI
observations the source showed a dramatic decrease in its count rate over the
years (\cite{dgrupe-WB1:gr01a}).  
The X-ray outburst during the RASS was 
followed by a response in its optical emission lines.
While in the May 1991 spectrum (Fig. \ref{dgrupe_WB1:ic3599_opt}; 
\cite{dgrupe-WB1:bra95}) the H$\beta$ line
as well as highly ionized Fe line (e.g. [FeX]) appear to be extremely strong,
in the spectra taken years later (Fig. \ref{dgrupe_WB1:ic3599_opt}; 
\cite{dgrupe-WB1:gr95a}) these lines have
become much weaker, but other lines (e.g. FeVII) showed up instead. 
A possible explanation of this X-ray outburst is an accretion
event either caused
by an instability of the accretion disk or by a tidal disruption of
a star orbiting around the central black hole. Such tidal disruption events have
been proposed by e.g. \cite{dgrupe-WB1:ree90}. They are a consequence if every
galaxy habours a supermassive black hole in its center (see
\cite{dgrupe-WB1:ree89}).

\subsection{RX J1624.9+7554}
RX J1624.9+7554 has shown a dramatic decrease of its X-ray flux by at
least a factor of more than 200 between the RASS and a pointed observation 1.5
years later (\cite{dgrupe-WB1:gru99}). 
Optical spectroscopy performed 8 years after the RASS, identified this 
source as a normal non-active galaxy. Because of the lack of any signs of
nuclear activity in this galaxy,
the most plausible explanation for this X-ray event
is the tidal disruption of a star by the central black hole. With a 0.2-2.0 keV
luminosity $L_{\rm X}$=1.6 10$^{37}$ W during the outburst it was the 
brightest non-active transient observed so far.

\subsection{RX J2217.9--5941}
The NLS1 
RX J2217.9--5941 is a possible X-ray transient candidate. It is highly variable on
time scales of days as well as years (\cite{dgrupe-WB1:gr01b}).
During its two-day
RASS observation the count rate decrease by a factor of 15. Observed several
times in pointed observations by ROSAT and ASCA the source has become fainter
over the years.
It is not clear yet if this source will become a transient. However, due to the
black hole mass of $\approx$ 10$^8$\msun the timescales are larger than in e.g.
IC 3599 (M$_{BH}\approx 10^6$\msun). RX J2217.9--5941 might be a bigger brother
of WPVS007 and we observe RX J2217.9--5941 slowly become fainter over the years.
They both have in common to be NLS1s and have steep X-ray spectra at least
during their RASS observation.

\subsection{Other X-ray transient galaxies}
Besides RX J1624.9+7554 a few more non-active galaxies have been reported of in
the literature: NGC 5905, RX J1242.6--1119, and RX J1420.4+5334
(\cite{dgrupe-WB1:ko99a}; \cite{dgrupe-WB1:ko99b}; \cite{dgrupe-WB1:gre00}).
While NGC 5905 was observed in an outburst state during the RASS and a later
turn-off, the two other
sources are the only examples of a `turn-on'. Both where brighter when observed
in pointed observations after the RASS. All the three sources have in common not
to show any signs of nuclear activity. The most plausible explanation for the
outburst is like in RX J1624.9+7554 a tidal disruption of a star by the central
black hole. 

\section{Discussion}
The ROSAT All-Sky Survey has shown the potential of soft X-ray surveys
to discover X-ray transient AGN and galaxies. 
There are in principle two types of transience in AGN and galaxies: a) a sudden
decrease of the X-ray flux in normally bright X-ray sources, and b) an X-ray
outburst caused by either an accretion disk instability or by a tidal disruption
of a star orbiting the central black hole. The first case, which might have been
seen in WPVS 007 and RX J2217.9--5941, might be the transition between a high
into a low state. Such behaviour is known from galactic black hole candidates.
The second case, an X-ray outburst, was observed in IC 3599, NGC 5905, RX
J1624.9+7554, RX J1242.6--1119, and RX J1420.4+5334.
The follow-up optical
observations of IC
3599 showed how the light front moved through the inner region of the AGN/galaxy
and
caused different emission lines to show up over time. While reverberation
mapping is a powerful tool in bright, `normal' AGN to map the inner region,
X-ray
transient AGN and galaxies show much larger and dramatic changes in their
optical spectra. The only problem is, such X-ray outburst events are rare.
In order to find those extraordinary sources, soft X-ray surveys are needed.

New soft X-ray surveys will provide us with a statistically relevant data set
that can clarify how common the chance between low and high states is in AGN.
Repeating soft X-ray surveys will also give us an estimate how often outburst
events appear in non-active galaxies, or in other words, how many years we have
to wait until such an event happens in a galaxy. This result is also of interest
for our own galaxy, which black hole mass of 2.6 10$^6$\msun is comparable to
the ones found in the galaxies in which an outburst occurred. 

The requirements for
future soft X-ray surveys are being sensitive at energies $<$0.3 keV and 
using an imaging X-ray telescope with a high effective area and a large field 
of view. A sensitivity in soft X-rays ($<$0.3 keV, at least) is needed, because
X-ray outbursts have a very soft X-ray spectrum, due to the high accretion rate
during the event. A high effective area is needed for getting enough photons to
derive spectra and lightcurves on short timescales. And last but not least, a
large field of view is needed to get a long coverage of the source when passing
through the field of view of the telescope.  

Currently, three X-ray surveys are planned, ROSITA, Lobster Eye, and MAXI. 
They all have in common to work at higher energies. While ROSITA  may
have enough spatial resolution to identify the optical counterpart of an X-ray
transient, Lobster Eye is an all-sky monitor which does not have enough spatial
resolving power. MAXI will not have enough spatial resolution either, plus will
not be sensitive to low energy photons.
ROSITA will be a powerful tool to discover obscured hard AGN,
but inefficient for the search of short living soft X-ray events such as X-ray
outburst. The other problem is that the number of photons collected per scan is
too small to derive lightcurves and spectra. Therefore, none of the currently
planned X-ray survey missions fullfill the requirements for a search for soft 
X-ray transient AGN and galaxies. Therefore a new soft X-ray survey mission is
needed, performed in a similar way as the RASS was, except it should be repeated
several times in order to a) get long-term light curves, b) see the same number
of `turn-ons' as `turn-offs'. Such a survey will not be only important for
AGN/galaxy research, it is also needed for the discovery and study of e.g. new
cataclysmic variables and super-soft X-ray sources.  

\begin{acknowledgements}
I would like to thank Dr. Stefanie Komossa for carefully reading the manuscript.
The ROSAT project was supported by the Bundesministerium f\"ur Bildung
und  Forschung (BMBF/DLR) and the Max-Planck-Society (MPG).
\end{acknowledgements}

\end{document}